\documentclass{article}
\pdfoutput=1

\usepackage{amssymb,amsfonts,amsmath}
\usepackage{cite,enumerate,float}
\usepackage{color}
\usepackage{tikz}
\usetikzlibrary{arrows,snakes,backgrounds}

\def\be{\begin{eqnarray}}
\def\ee{\end{eqnarray}}
\def\nn{\nonumber}

\def\p{\partial}
\def\tr{{\rm tr}\,}
\def\Tr{{\rm Tr}\,}

\definecolor{red}{rgb}{1,0,0}
\definecolor{orange}{rgb}{1,0.5,0}
\definecolor{violet}{rgb}{0.7,0,1}

\def\cre{\color{red}}

\def\cb{\color{blue}}

\hoffset= - 1.0in         

\begin{document}

\begin{center}
\begin{small}
\hfill MIPT/TH-18/22\\
\hfill FIAN/TD-13/22\\
\hfill ITEP/TH-21/22\\
\hfill IITP/TH-20/22\\

\end{small}
\end{center}

\vspace{.5cm}

\begin{center}
\begin{Large}\fontfamily{cmss}
\fontsize{15pt}{27pt}
\selectfont
	\textbf{Spectral curves and $W$-representations of matrix models}
	\end{Large}
	
\bigskip \bigskip

\begin{large}
A. Mironov$^{b,c,d}$\,\footnote{mironov@lpi.ru; mironov@itep.ru},
A. Morozov$^{a,c,d}$\,\footnote{morozov@itep.ru}
\end{large}

\bigskip

\begin{small}
$^a$ {\it MIPT, Dolgoprudny, 141701, Russia}\\
$^b$ {\it Lebedev Physics Institute, Moscow 119991, Russia}\\
$^c$ {\it Institute for Information Transmission Problems, Moscow 127994, Russia}\\
$^d$ {\it NRC ``Kurchatov Institute'' - ITEP, Moscow 117218, Russia}\\
\end{small}
 \end{center}

\bigskip

\begin{abstract}
We explain how the spectral curve can be extracted from the
${\cal W}$-representation of a matrix model.
It emerges from the  part of the ${\cal W}$-operator,
which is linear in time-variables.
A possibility of extracting the spectral curve in this way
is important because there are models where matrix integrals
are not yet available, and still they possess all their
important features.
We apply this reasoning to the family of WLZZ models
and discuss additional peculiarities which appear for the
non-negative value of the family parameter $n$,
when the model depends on additional couplings (dual times).
In this case, the relation between topological and $1/N$ expansions
is broken. On the other hand, all the WLZZ partition functions are
$\tau$-functions of the Toda lattice hierarchy, and these models also
celebrate the superintegrability properties.
\end{abstract}

\bigskip

\section{Introduction}

Many properties of matrix models \cite{Mehta,UFN3} are defined by
their spectral curves,
which define the distribution of eigenvalues in the large $N$ limit,
and is a generating function of all the genus zero contributions
to the single-trace correlators.
This does not seem to be much,
still, if the whole set of Virasoro-like constrains
is available, the spectral curve (with some simple additional data) is sufficient
to reproduce all correlators at all genera,
the relevant procedure is known as the AMM-EO topological recursion \cite{AMMtr,EO}.
In this sense, the knowledge of the spectral curve is nearly equivalent
to that of the entire matrix model.

On the other hand, nowaday matrix model partition functions are defined not only
by an explicit matrix (or eigenvalue) integral, but also by action of
an operator $\hat {\cal W}$ on a trivial state in the space of matrix model couplings $p_k$:
\be
Z\{p\} = e^{\hat {\cal W}\{p\}}\cdot 1
\label{wrep}
\ee
Such a realization is called $W$-representation \cite{Wrep,Alex,Max1,Max,MMMR,MMM} (see also similar realizations in \cite{Giv,Ok,AMM,AMMtr,BM,MMkhur,Kaz}).
Sometimes, it is better to present it in the form
\be\label{wrep2}
Z\{p\} = e^{\hat {\cal W}\{p\}}\cdot e^{\sum_k g_kp_k/k}
\ee
where $g_k$ are parameters.
Such second form can be definitely reduced to (\ref{wrep}) using the Campbell-Hausdorff formula, but the resulting $\hat {\cal W}$-operator is too complicated. Hence, the form (\ref{wrep2}) is more preferable in such a case.

It is a natural question to ask what is the spectral curve, and the topological recursion
in terms of this $W$-representation.
Once understood, this would provide spectral curves
for models that are not defined through any integrals.

Our claim in this paper is that the spectral curve is associated
with a peculiar part $\hat {\cal W}^{\rm spec}$ of the $\hat {\cal W}$-operator.
We demonstrate this in detail for the Gaussian model,
and then discuss implication for the other cases, mostly for the WLZZ models \cite{China}. In particular, we give a general recipe for constructing $\hat {\cal W}^{\rm spec}$ in these cases.
More formally, our claims are:
\begin{itemize}
\item $\hat{\cal W}^{\rm spec}$ is made from all the terms of $\hat{\cal W}$,
which are linear in $p$
\\ (but these terms can be non-linear in $p$-derivatives, like
$\frac{\p^2}{\p p^2}$ in the Gaussian model)

\item The action of $e^{\hat{\cal W}^{\rm spec}}$ produces an exponential of an expression,
which is {\bf linear in $p$}
\\ (and no longer contains $p$-derivatives)
\be
e^{\hat {\cal W}^{\rm spec}}\cdot 1 = \exp\left({\cal P}\right)
\ee
Thus we see {\bf a mysterious role of the exponential function}.

\item Making the substitution $p_k\to z^kp_k$ allows one to generate the function (resolvent) $y(z)$ such that
\be
{\cal P}(z)=\oint V(xz)\,y(x)  dx
\ee
where $V(z)=\sum_kp_kz^k/k$ is the matrix model potential. The resolvent $y(z)$ satisfies the spectral curve equation.

\item For WLZZ models \cite{China} with negative grading $m<0$, the resolvent
\be
y_m(z) = \sum_{k} \frac{|m|k}{z^{|m|k+1}} \frac{\p \log {\cal Z} }{\p p_{|m|k}}
\ee
satisfies the spectral curve that is a simple generalization of the semicircle distribution equation at $m=-2$:
\be
y_m^{|m|}  - zy_m +1 = 0
\ee
This spectral curve equation describes the large $N$ limit, and corresponds to the leading behaviour of the topological expansion. The complete expansion is constructed from the full set of $W$-constraints
\be
\left(n\frac{\p}{\p p_{n}} +\widetilde {W}_{n-m}^{(-,m)}\right) Z = 0
\ee
where the operators $\hat {\cal W}_{-m}$ generating the $W$-representation are connected with $\widetilde {W}_{n-m}^{(-,m)}$ by the relation
\be
\hat{\cal W}_{-m} = \sum_{k=1}^\infty  p_k \widetilde {W}_{n-m}^{(-,m)}
\ee

\item In the WLZZ model at $m=0$, the ${\cal W}_0$ annihilates $1$, and one should act on a non-nontrivial
state $\exp\left(\beta p_1\right)$. The spectral curve is given by the Lambert curve. One can naturally extend this ${\cal W}_0$-operator to a series of operators associated with the generalized cut-and-join operators $W_{[s]}$, which gives rise to higher Lambert curves (the ${\cal W}_0$ case corresponds to $s=2$)
\be
ye^{-z^{s-1}y^{s-1}}={\beta\over z^2}
\ee

\item In the WLZZ model at $m> 0$, the  ${\cal W}_m$ also annihilates $1$,
but one should act on a non-nontrivial
state $\exp\left(\sum_k g_kp_k\right)$.
This makes the situation more involved and intriguing, because one acquires new
parameters (dual time-variables) $g_k$.
In this case, the spectral curve is given by the equation
\be
y=\sum_{k=2}{g_k\over z^{k+1}}\Big(1+z^{2m-1}y^{m-1}\Big)^{k-1}
\ee
and is associated with the {\bf small}(!) $N$ expansion,
so that the interpretation in terms of topological expansion is no longer straightforward.

\item Superintegrability relations trivialize in the large $N$ limit:
the averages become {\it linear} in each sector of a given grading
\be
\Big< S_R\Big>_\infty = N^{|R|} S_R\{\delta_{k,n}\}
\ee
and multiple correlators of characters factorize
\be
\Big< \prod_i S_{R_i}\Big>_\infty =\prod_i \Big< S_{R_i}\Big>_\infty
\ee
where $S_R$ are the Schur functions \cite{Mac}.
The Schur functions can be treated as symmetric polynomials of variables $x_i$, or graded polynomials
of power sums $p_k=\sum_ix_i^k$. In the later case, we use the notation
$S_R\{p_k\}$.
We also use the shortened notation $S_R\{N\}=S_R\{p_k=N\}$ and $S_R\{\delta_{k,m}\}=S_R\{p_k=\delta_{k,m}\}$.
\end{itemize}

In the paper, we are mostly considering the WLZZ partition functions \cite{China},
which are introduced by $W$-representations:
\be
Z_m = e^{\hat {\cal W}_m/m} \cdot
\left\{\begin{array}{ccc} 1 & {\rm for} & m<0 \\ \\ \exp\left(\sum_k  \frac{g_kp_k}{k}\right)
  & {\rm for} & m>0
\end{array}\right.
\ee
and the Hurwitz partition function \cite{Ok,MMN}
\be
Z_0=e^{\hat {\cal W}_0}\cdot e^{\beta p_1}
\ee
These definitions are inverse to the ordinary definition, when one searches for a ${\cal W}$-representation
for a given model with nice properties.
There is no {\it a priori} reason to expect that this new formulation is {\it obliged}
to provide interesting results.
However, we demonstrate that it does. The results look relatively simple for the negative branch of the WLZZ models.
In this case, the WLZZ models are described by the two-matrix models of the Hurwitz type \cite{AMMN2,Al,Ch}
counting the isomorphism classes of Belyi pairs,
arising in the study of equilateral triangulations
and  Grothiendicks's {\it dessins d'enfant} \cite{Gr}.

Things get more intriguing for $m=0$, related both with the Hurwitz numbers and, by the ELSV formula, with the
Kontsevich model deformation involving the Hodge integrals \cite{BM,MMkhur},
and even more intriguing for positive $m$, when the WLZZ models are described by two-matrix models in an external field \cite{Ch}.

The WLZZ family includes two one-matrix model examples: for $m=\pm 2$. Another example well-studied earlier is $m=0$.
We underline the subsections devoted to these three particular cases.

Let us stress that our definition of the spectral curve follows the recursion procedure worked out in \cite{AMM}: one starts from the loop equation for the resolvent and than considers the leading order that excludes terms with derivatives of the resolvent. In this leading order, the loop equation becomes an algebraic equation, which is the spectral curve. The recursion is further constructed as an expansion of the resolvent over this spectral curve. In this paper, we directly applied this procedure in the cases of $m=\pm 2$ for the corresponding one-matrix models, and managed to confirm our scheme for constructing the spectral curve (in particular, the expansion parameters of large and small $N$ for $m=2$ and $m=-2$, correspondingly).

Finally, let us note that manifest expressions for the results of action of the $W$-operators like (\ref{47}) in this paper are conjectural and checked by computer evaluations of many terms of expansion of the exponential and at many values of $m$ (both negative and positive). One can prove such formulas in simple cases like (\ref{25}), (\ref{42}) reducing them to an equation like (\ref{31}), however, in other, more complicated cases this way of proof becomes already too involved.

\section{\underline{Basic example: Hermitian Gaussian model}}

\subsection{Description of the model}
In this section, we explain how one can construct the spectral curve from the $W$-representation in the simplest case of the Hermitian Gaussian matrix model. The partition function of this model satisfies an infinite set of Virasoro constraints (generated by the Borel  subalgebra of the Virasoro algebra):
\be\label{22}
L_nZ_{-2}=(n+2){\p Z_{-2}\over\p p_{n+2}},\ \ \ \ \ n\ge -1\nn\\
L_n=\sum_{k} (k+n)p_k\frac{\p}{\p p_{k+n}} + \sum_{a=1}^{n-1} a(n-a)\frac{\p^2}{\p p_a\p p_{n-a}}
+ 2Nn\frac{\p}{\p p_n}    + N^2\delta_{n,0} + Np_1 \delta_{n+1,0}
\ee
and has a ${\cal W}$-representation of the form
\be\label{W2}
Z_{-2}=e^{{1\over 2}\hat {\cal W}_{-2}}\cdot 1
\ee
\be
\hat {\cal W}_{-2}=\sum_kp_k\hat L_{k-2}=
\sum_k(k+l-2)p_kp_l{\p\over\p p_{k+l-2}}+\sum_{k,l}klp_{k+l+2}{\p^2\over\p p_k\p p_l}+2N\sum_kkp_{k+2}{\p\over
\p p_k}+N^2p_2+Np_1^2\nn
\ee
Because the r.h.s. of the Virasoro constraints is of different grading,
this partition function is an unambiguous solution to the non-trivial equation
\be
\Big(\hat l_0-\hat {\cal W}_{-2}\Big)Z_{-2}=0,\ \ \ \ \ \hat l_0:=\sum_k kp_k{\p\over\p p_k}
\ee
This solution is actually equal to
\be
Z_{-2}=\sum_R{S_R\{N\}S_R\{\delta_{k,2}\}\over d_R}S_R\{p_k\}
\ee
which is a sum over the Schur functions with factorized coefficients proportional to the same Schur functions evaluated at special loci. Such specific form of a sum is nicknamed a {\it superintegrability} property \cite{siMM,MMsirev,MMZU}.

\subsection{Spectral curve
\label{ssc}}

Let us now make the following trick \cite{AMM}:

\bigskip

1) introduce the variables $t_k:=\frac{p_k}{k\hbar}$,

2) introduce $t_0$ such that ${\p Z_{-2}\over\p t_0}:=NZ_{-2}$.

\bigskip

\noindent
Then, one can rewrite the Virasoro generators (\ref{22}) in the form
\be\label{Vircon}
L_n=\sum_{k=0}kt_k{\p\over\p t_{k+n}}+\hbar^2\sum_{k\ge 0}{\p^2\over\p t_k\p t_{n-k}}
\ee
Let us now define the resolvent
\be
\rho_{-2}(z|t_k):=\hat\nabla_z{\cal F}_{-2}=\sum_{k\ge 0}{1\over z^{k+1}}{\p {\cal F}_{-2}\over\p t_k}\nn\\
{\cal F}_{-2}:=\hbar^2\log Z_{-2}
\ee
Then, the generating function of all the Virasoro constraints by converting $L_n$'s with $z^{-n-2}$ can be rewritten in the form:
\be\label{leq}
\rho_{-2}(z|t_k)^2-z\rho_{-2}(z|t_k)+N+\hbar^2\hat\nabla_z\rho_{-2}(z|t_k)+\Big[V'(z)\rho_{-2}(z|t_k)\Big]_-
\ee
where \ $V(z):=\sum_{k=1} \frac{p_{k}}{k} z^{k}=\sum_{k=1} t_{k} z^{k}$ \ \
is the potential of the matrix model, and $\Big[\ldots\Big]_-$ denotes projection to the negative powers of $z$.

Consider now the solution at the leading (spherical) order at small $\hbar$ and at zero $p_k$ at all $k$, $y:=\rho_{-2}(z|0)\Big|_{\hbar\to 0}$. Then, one obtains the equation
\be\label{sc2}\boxed{
y^2-zy+N=0}
\ee
which is exactly the spectral curve.
Actually, in this particular case this the Riemann sphere in hyperelliptic representation.

Parameter $\hbar$ can be used to define topological expansion \cite{TE,AMM}
and AMM-EO topological recursion \cite{AMMtr,EO}.
In this case, one can identify  $\hbar=1/N$ by rescaling time-variables $t_k\longrightarrow Nt_k$
(i.e. by making the substitution $\log Z_{-2} = N^2{\cal F}_{-2}$):
this allows one {\bf to identify topological and $1/N$ expansions}.
As we shall see, this identification appears consistent for all WLZZ models with $m<0$,
but breaks down for $m\geq 0$.

\subsection{Spectral curve from the $W$-representation}

Now let us note that the leading order in the loop equations is completely due to
the second derivatives terms in $\hat L_n$ and, hence, in $\hat{\cal W}_{-2}$.
Hence, one could naturally expect that, in order to generate the leading term ${\cal F}_0$,
one has to truncate the ${\cal W}$-representation to the second derivative terms only:
\be
\hat {\cal W}_{-2}^{\rm spec} :=
\sum_{a,b\geq 0} (a+b+2)t_{a+b+2}\frac{\p^2}{\p t_a\p t_b}
= \sum_{a,b\geq 1}
 abp_{a+b+2}\frac{\p^2}{\p p_a\p p_b} + 2N\sum_{a\geq 1} ap_{a+2}\frac{\p}{\p p_a} + N^2p_2
\ee
Now we define
\be
e^{\frac{z^2}{2} \hat {\cal W}_{-2}^{\rm spec}} \cdot 1 = e^{{\cal P}_{-2}(z)}
\label{W2largeN}
\ee
and realize that
\be\label{25}
{\cal P}_{-2}(z)= N\sum_{k=1}^\infty \frac{p_{2k}}{2k}
\frac{2\cdot\Gamma(2k)}{\Gamma\big(k+2\big)\Gamma(k)}(Nz^2)^k
\ee
 Note that the exponential ${\cal P}$ is linear in $p_k$.  This {\bf exponentiation} phenomenon in (\ref{W2largeN}) takes place for all other models in this text. It is not a priori that obvious, and it is explained by the structure of the Campbell-Hausdorff formula, see s.\ref{CH}.

On the other hand, expanding (\ref{sc2}) at large $z$, one obtains
\be
y(z) = \frac{z-\sqrt{z^2-4N}}{2}=y(z) = \frac{N}{z}+  \sum_{k=1}^\infty  \frac{2\cdot\Gamma(2k)}{\Gamma\big(k+2\big)\Gamma(k)}
\cdot z^{-2k-1}N^{k+1}
\label{rhoGauss}
\ee
Thus, one obtains
\be
{\cal P}_{-2}(z)=  \oint V(xz)\,y(x)  dx
\ee
As we already pointed out, one expects that ${\cal P}_{-2}(z)$ is a leading contribution to the partition function at small $g$. Indeed, this formula can be compared with \cite[Eq.(48)]{ChMMV} to see this is really the case. Here we derived this contribution completely in terms of the $W$-representation.

Note that the $N$-dependence for $\hat{\cal W}^{\rm spec}$
can be fully eliminated by the change of variables $p_k\longrightarrow p_k/N$
and $z^2\longrightarrow z^2/N$.
Remarkably, this is a kind of opposite to the change $p_k\longrightarrow Np_k$, relevant for the
definition of the $1/N$ expansion at the end of the section \ref{ssc}.

Note that the true partition function corresponds to the particular value $z^2=1$ in
(\ref{W2largeN}) with the full-fledged operator $\hat{\cal W}_{-2}$.
Spectral curve, however, appears when we truncate the operator to $\hat{\cal W}^{\rm spec}_{-2}$
and {\it release} $z$. As we explain in secs.4-5, this procedure can be formulated in another, more
universal way by making a substitution $p_k\to z^kp_k$ instead of releasing $z$.

\section{An infinite set of WLZZ models. Negative branch\label{neg}}

\subsection{Description of the models\label{negd}}

The authors of \cite{China} proposed an infinite set of models parameterized by integers. The models parameterized by negative integers
generalize the Hermitian Gaussian model and are described by the following procedure:
one constructs the ${\cal W}$-representations of these models using the operators built by a recursive procedure
\be\label{Wn}
\hat {\cal W}_{-m-1}={1\over m!}[\hat {\cal W}_{-1},\hat {\cal W}_{-m}],\ \ \ \ \ m\ge 2
\ee
where
\be
\hat {\cal W}_{-1}:=\sum_k(k+l-1)p_kp_l{\p\over\p p_{k+l-1}}+\sum_{k,l}klp_{k+l+1}{\p^2\over\p p_k\p p_l}+2N\sum_kkp_{k+1}{\p\over
\p p_k}+N^2p_1
\ee
Every such operator gives rise to a partition function
\be\label{Zmn}
Z_{-m}=e^{{1\over m}\hat {\cal W}_{-m}}\cdot 1
\ee
which is an unambiguous solution to the equation
\be\label{31}
\Big(\hat l_0-m \hat {\cal W}_{-m}\Big)Z_{-n}=0,\ \ \ \ \ \hat l_0:=\sum_k kp_k{\p\over\p p_k}
\ee
Among other things, this means that the matrix models that describes $Z_{-m}$ could not be of the $X^m$-potential type: there should be only one possible integration contour. This is, indeed, the case, see (\ref{mamo}).

The partition function, (\ref{Zmn}) is equal (for $m\ge 2$) to
\be\label{WLZZnsi}
Z_{-m}=\sum_R{S_R\{N\}S_R\{\delta_{k,m}\}\over d_R}S_R\{p_k\}
\ee
and is a KP $\tau$-function of the hypergeometric type \cite{GKM2,OS,AMMN1,AMMN2}.

Representation (\ref{WLZZnsi}) for the partition function implies the superintegrability relation for the correlator \cite{MMsirev},
\be\label{sin}
\Big<S_R\{P_k\}\Big>={S_R\{N\}S_R\{\delta_{k,m}\}\over S_R\{\delta_{k,1}\}}
\ee
where $P_k=\tr M^k$ are traces of matrices at the matrix WLZZ model.

This partition function (\ref{WLZZnsi}) can be realized by the two-matrix integral \cite{AMMN2,Al}
\be\label{mamo}
Z_{-m}=\int\int_{N\times N}dXdY\exp\left(-\Tr XY+\sum_k {p_k\over k}\Tr X^k+{1\over m}\Tr Y^m\right)
\ee
Here the integral is understood as integration of a power series in $p_k$, and $X$ are Hermitian $N\times N$ matrices, while $Y$ are anti-Hermitian $N\times N$ ones. Such partition function satisfies the so called $\widetilde W$-algebra constraints \cite{Gava}:
\be\label{Wmcon}
\widetilde {W}_{n}^{(-,m)}Z_{-m}=(n+m){\p Z_{-m}\over\p p_{n+m}}, \ \ \ \ \ n\ge -m+1
\ee
There are spin $m$ $\widetilde W$-algebras of two types which we denote $\widetilde W^{(\pm,m)}$, their generators being defined \cite{Gava,GKMU} (see also \cite[sec.7]{AMMN2}) by any of the following three relations\footnote{Hereafter, by the matrix derivative, we imply the derivative w.r.t. matrix elements of the transposed matrix: $\left(\frac{\partial}{\partial \Lambda}\right)_{ij}=\frac{\partial}{\partial \Lambda_{ji}}$.}:
\be
\left.
\left(\frac{\partial}{\partial \Lambda}\right)^{m+1} f(p_k) =
\sum_{n\geq 1} \Lambda^{\pm n -1}\widetilde { W}_{n \pm m}^{(\pm,m+1)}(p_k)
f(p_k)
\right|_{p_k =  { \Tr} \Lambda^{\pm k}}
\label{W-tilde-1}
\ee
or\footnote{In eq.(\ref{W-tilde-2}), we again introduced $t_k:=p_k/k$ in order to switch on the variable $t_0$,
which makes the formulas simpler.}
\be
\widetilde {W}_{n \pm m}^{(\pm,m+1)}(t)
e^{\sum_{k\geq 0} t_k{\Tr}\Lambda^{\mp k}} =
{\Tr}\left\{\left(\frac{\partial}{\partial \Lambda}\right)^m
\Lambda^{\mp n}\right\}e^{\sum_{k\geq 0} t_k{\Tr}\Lambda^{\mp k}}
\label{W-tilde-2}
\ee
or
\be
\widetilde{W}_{n }^{(\pm,m+1)}(p) = \sum_{k\geq 1} p_k
\widetilde{W}_{n + k}^{(\pm,m)}(p) +
\sum_{k=1}^{n\mp (m-1/2)-1/2}k\frac{\partial}{\partial p_k}
\widetilde{W}_{n-k }^{(\pm,m)}(p)+N\widetilde{W}_{n}^{(\pm,m-1)}
\label{W-tilde-rec}
\ee
The last recurrence relation should be supplemented by ``the initial
condition''
\be\label{WI1}
\begin{array}{c}
\widetilde{W}^{(\pm,1)}_n = n\frac{\partial}{\partial p_n},\ \ n\geq 1\cr
\cr
\widetilde{W}^{(\pm,1)}_0 = N\cr
\end{array}
\ee
and one generally requires that
\be\label{rec}
\widetilde{W}^{(\pm,m)}_n = 0,\ \ n\leq -m
\ee

\subsection{Model with $m\geq 3$}

\subsubsection{${\cal W}$-representation of the $m=3$ model}

Let us start from the simplest example of $m=3$. Then, one can obtain from (\ref{Wn}) and (\ref{W2})
\be
\hat {\cal W}_{-3}={1\over 2}[\hat {\cal W}_{-1},\hat {\cal W}_{-2}]=\sum_{k,l,m}mp_kp_lp_{m-k-l+3}{\p\over\p p_m}+
{3\over 2}\sum_{k,l,m}kmp_lp_{k+m-l+3}{\p^2\over\p p_m\p p_k}+\nn\\
+\sum_{k,l,m}klmp_{k+l+m+3}{\p^3\over\p p_k\p p_l\p p_m}
+{1\over 2}\sum_kk(k+1)(k+2)p_{k+3}{\p\over\p p_k}+\nn\\
+3N\sum_{k,l}klp_{k+l+3}{\p^2\over\p p_k\p p_l}+3N\sum_{k,l}lp_kp_{l-k+3}{\p\over\p p_{l}}
+N(p_1^3+p_3)+\nn\\
+2N^2\sum_kkp_{k+3}{\p\over\p p_k}
+3N^2p_1p_2+N^3p_3
\ee
In accordance with the general rule of constructing ${\cal W}$-representations \cite{MMMR}, this operator can be rewritten as
\be\label{sum}
\hat {\cal W}_{-3}=\sum_kp_k\hat W^{(3)}_{k-3}
\ee
with a simple and immediate operator
\be\label{W3}
\hat W^{(3)}_n=\sum_{m,l}(n+m+l)p_lp_m{\p\over\p p_{m+l+n}}+
{3\over 2}\sum_{m,l}lmp_{l+m-n}{\p^2\over\p p_l\p p_m}+\nn\\
+\sum_{m,l}(n-m-l)ml{\p^3\over\p p_m\p p_l\p p_{n-m-l}}+{n(n+1)(n+2)\over 2}{\p\over\p p_n}+\nn\\
+3N\sum_k(k+n)p_k{\p\over\p p_{k+n}}+3N\sum_kk(n-k){\p^2\over\p p_k\p p_{n-k}}+3N^2n{\p\over\p p_n}+\nn\\
+N(N^2+1)\delta_{n,0}
+3N^2p_1\delta_{n,-1}+Np_1^2\delta_{n,-2}
\label{W3op}
\ee
One may think that again, similarly to the $m=-2$ case, the infinite set of constraints satisfied by the partition function is
\be\label{W3f}
\hat {W}_{n}^{(3)}Z_{-3}\stackrel{?}{=}(n+3){\p Z_{-3}\over\p p_{n+3}}, \ \ \ \ \ n\ge -2
\ee
However, this is not the case: for instance,
\be
\hat {W}_{-2}^{(3)}Z_{-3}-{\p Z_{-3}\over\p p_{1}}=-2N^2p_2+\ldots\nn\\
\hat {W}_{-1}^{(3)}Z_{-3}-2{\p Z_{-3}\over\p p_{2}}=2N^2p_1+\ldots
\ee
and only in the sum (\ref{sum}) the first terms cancel with each other. It is related to the fact is that the set of equations (\ref{W3f}) does not admit any non-trivial solutions.

The point is that the representation of $\hat {\cal W}_{-3}$ as a sum (\ref{sum}) is not unique, and
one can equivalently realize $\hat {\cal W}_{-3}$ instead of (\ref{sum}) as
\be\label{sum1}
\hat {\cal W}_{-3}=\sum_kp_k\widetilde {W}_{k-3}^{(-,3)}
\ee
with
\be\label{WI3}
\widetilde{W}^{(-,3)}_n=\sum_{a+b+c=n}abc{\p^3\over\p p_a\p p_b\p p_c}
+\sum_{k=1}p_k\left(\sum_{a=1}^{n+1}+\sum_{a=1}^{n+k-1}\right)a(k+n-a){\p^2\over\p p_a\p p_{k+n-a}}+\nn\\
+\sum_{a,b=1}(a+b+n)p_ap_b{\p\over\p p_{a+b+n}}+{n(n+1)(n+2)\over 2}{\p\over\p p_n}+\nn\\
+3N\sum_{a=1}^{n-1}a(n-a){\p^2\over\p p_a\p p_{n-a}}+3N\sum_{k=1}(n+k)p_k{\p\over\p p_{n+k}}
+3N^2n{\p\over\p p_n}+N(n+1)p_1{\p\over\p p_{n+1}}+\nn\\
+N(N^2+1)\delta_{n,0}+2N^2p_1\delta_{n,-1}+N(p_1^2+Np_2)\delta_{n,-2}-N\sum_kkp_{k+2}{\p\over\p p_k}\delta_{n,-2}
\ee
and the equation
\be\label{W3con}
\widetilde {W}_{n}^{(-,3)}Z_{-3}=(n+3){\p Z_{-3}\over\p p_{n+3}}, \ \ \ \ \ n\ge -2
\ee
 is satisfied, in accordance with (\ref{Wmcon}).

\subsubsection{Spectral curve}

In order to construct the loop equation, one has to repeat the procedure of sec.\ref{ssc}. The only difference is that now one has to convert the constraints (\ref{W3con}) at $m=3$ with $z^{-n-3}$. Again,
\be
\rho_{-3}(z|t_k):=\hat\nabla_z{\cal F}_{-3}=\sum_{k\ge 0}{1\over z^{k+1}}{\p {\cal F}_{-3}\over\p t_k}
\ee
and introducing $y:=\rho_{-3}(z|0)\Big|_{g\to 0}$, one obtains the spectral curve equation
\be\label{sc3}\boxed{
y^3-zy+N=0}
\ee

\subsubsection{Spectral curve from  the ${\cal W}$-representation}

In order to relate solution of the spectral curve equation, (\ref{sc3}) with the ${\cal W}$-representation, we again
truncate the ${\cal W}$-representation to the third derivative terms only:
\be
\hat {\cal W}_{-3}^{\rm spec} :=
\sum_{a,b,c\geq 0} (a+b+c+3)t_{a+b+c+3}\frac{\p^3}{\p t_a\p t_b\p t_c}=\nn\\
= N^3p_3+3N^2\sum_{a=1}^\infty ap_{a+3}\frac{\p}{\p p_a}
+3N\sum_{a,b=1}^\infty abp_{a+b+3}\frac{\p^2}{\p p_a\p p_b}
+ \sum_{a,b,c=1}^\infty abcp_{a+b+c+3}\frac{\p^3}{\p p_a\p p_b\p p_c}
\ee
We again define
\be
e^{\frac{z^3}{3} \hat {\cal W}_{-3}^{\rm spec}} \cdot 1 = e^{{\cal P}_{-3}(z)}
\ee
and realize that
\be\label{42}
{\cal P}_{-3}(z)= N\sum_{k=1}^\infty  \frac{3\Gamma(3k)(z^3 N^2)^k}{\Gamma(2k+2)\Gamma(k)}  \frac{p_{3k}}{3k}
\ee
On the other hand, expanding (\ref{sc3}) at large $z$, one obtains
\be
y(z) = \frac{N}{z}+  \sum_{k=1}^\infty  \frac{3\cdot\Gamma(3k)}{\Gamma\big(2k+2\big)\Gamma(k)}
\cdot z^{-3k-1}N^{2k+1}
\ee
Thus, one obtains
\be
{\cal P}_{-3}(z)=  \oint V(xz)\,y(x)  dx
\ee

\subsection{Model with generic $m$}

Formulas are just the same for any $\hat {\cal W}_{-m}^{\rm spec}$, for example
\be
\hat {\cal W}_{-4}^{\rm spec} :=
N^4p_4+4N^3\sum_{a=1}^\infty ap_{a+4}\frac{\p}{\p p_a}
+6N^2\sum_{a,b=1}^\infty abp_{a+b+4}\frac{\p^2}{\p p_a\p p_b}
+ \nn \\
+ 4N\sum_{a,b,c=1}^\infty abcp_{a+b+c+4}\frac{\p^3}{\p p_a\p p_b\p p_c}
+ \sum_{a,b,c,d=1}^\infty abcdp_{a+b+c+d+4}\frac{\p^4}{\p p_a\p p_b\p p_c\p p_d}
\ee
Defining again
\be
e^{\frac{z^m}{m} \hat {\cal W}_{-m}^{\rm spec}} \cdot 1 = e^{{\cal P}_{-m}(z)}
\ee
one obtains
\be\label{47}
{\cal P}_{-m}(z)= N\sum_{k=1}^\infty  \frac{m\Gamma(mk)(z^{m} N^{m-1})^k}{\Gamma((m-1)k+2)\Gamma(k)}  \frac{p_{mk}}{mk}
\ee
The spectral curve equation in this case is
\be\label{scpn}\boxed{
y^m - zy + N = 0}
\ee
Expanding its solution at large $z$, one obtains
\be
y(z) = \frac{N}{z}+  \sum_{k=1}^\infty  \frac{m\cdot\Gamma(mk)}{\Gamma\big((m-1)k+2\big)\Gamma(k)}
\cdot z^{-mk-1}N^{(m-1)k+1}
\ee
Thus, one again obtains
\be
{\cal P}_{-m}(z)=  \oint V(xz)\,y(x)  dx
\ee

One can  always invert the procedure and start from the ${\cal W}$-representation,
pick up the terms
with maximal number of derivatives, and calculate the corresponding ${\cal P}$. After this, one calculates $y(x)$ (up to the first term, which is always $N/z$)
and then find an equation that is satisfied by this $y$. This equation is exactly the spectral curve.
When only $p$-linear terms are kept, i.e. when we deal with $\hat{\cal W}^{\rm spec}$,
the matrix size $N$ can always be eliminated by the change
$p_k \longrightarrow p_k/N$, $z\longrightarrow z/N^{\frac{m-1}{m}}$. This scheme works for all the models we considered so far, however, it has to be improved in some points as we shall see in the next two sections.

\section{From Hurwitz model to Lambert curves}

\subsection{\underline{Hurwitz model and its spectral curve}
\label{00model}}

There is also a model in between negative and positive branches of the WLZZ models\footnote{The operator (\ref{W0}) of this model generates both the operator
$$
\hat {\cal W}_1=\Big[\hat{\cal W}_0,\Big[\hat{\cal W}_0,{\p\over\p p_1}\Big]\Big]
$$
and the operator
$$
\hat {\cal W}_{-1}=\Big[\hat{\cal W}_0,\Big[\hat{\cal W}_0,p_1\Big]\Big]
$$
generating the positive and negative branches accordingly, secs.\ref{pos} and \ref{neg}.}. This model
is given by the ${\cal W}$-representation
\be\label{W0}
\hat {\cal W}_0=\sum_{a,b}abp_{a+b}{\p^2\over\p p_a\p p_b}+\sum_{a,b}(a+b)p_ap_b{\p\over\p p_{a+b}}+N\sum_aap_a{\p\over\p p_a}
\ee
and is nothing but the Hurwitz partition function \cite{GJ,MMN}
\be
Z_0=e^{x\hat {\cal W}_0}\cdot e^{\beta p_1}=\sum_R\beta^{|R|}S_R\{\delta_{k,1}\}S_R\{p_k\}e^{xC_2(R)}
\ee
where $C_2(R)$ denotes the eigenvalue of second Casimir operator: $C_2(R)=\sum_{i,j\in R}(N+j-i)$. This partition function is also a KP $\tau$-function of the hypergeometric type \cite{GKM2,Ok,AMMN1,AMMN2}.

Now the spherical limit again is governed by the part of this operator with maximal number of derivatives:
\be
\hat {\cal W}_0^{\rm spec}=\sum_{a,b}abp_{a+b}{\p^2\over\p p_a\p p_b}+N\sum_aap_a{\p\over\p p_a}
\ee
One obtains
\be
e^{x \hat {\cal W}_0^{\rm spec}} \cdot e^{\beta p_1} = e^{{\cal P}_{0}(x)}
\ee
with
\be
{\cal P}_{0}(x)={1\over x}\sum_{k\ge 1}{(2k)^{k-1}\over k!}\Big(\beta e^{x N}x\Big)^k{p_k\over k}
\label{calP0}
\ee
This is the first time, when we can observe that the coefficient in front of the ${\cal W}$-operator, $x$ does not provide a good spectral parameter, since it no longer provides grading (because of the term $e^{xN}$). This is because the exponential of ${\cal W}$-operator acts not on the unity, but on the exponential of times. Hence, from now on, we use another procedure, which does not give anything new in the earlier considered cases, but will be of use in the forthcoming considerations. That is, we use $x$ as a free parameter that can be chosen in a convenient way (it can be removed by rescalings of other parameters), and instead we {\bf introduce the spectral parameter $z$ by making a substitution $p_k\to z^kp_k$}.

\bigskip

\begin{center}
\framebox{\parbox{12cm}{\bf Hence, our general prescription for making the spectral curve from the $ \hat {\cal W}^{\rm spec}$-operator is: \\ \\ (i) to use this operator instead of the full operator in the ${\cal W}$-representation; \\ \\
(ii) to demonstrate that this produces a linear exponential in $p_k$'s, and gives rise to ${\cal P}(z)$, where the $z$-dependence is introduced by making the substitution $p_k\to z^kp_k$; \\ \\
(iii) to use the formula
\be
{\cal P}_{0}(z)=  \oint V(wz)\,y(w)  dw\nn
\ee
in order to generate $y(z)$, which satisfies the spectral curve equation.
}}
\end{center}

\bigskip

In particular, in the model under consideration, we obtain from (\ref{calP0})
\be
y(z)={1\over 2x}\sum_{k\ge 1}{k^{k-1}\left(2x\beta e^{x N}\right)^k\over k!}z^{-k-1}
\ee
This is a large $z$ expansion of the spectral curve
\be
2x y e^{-2x yz}={\xi\over z^2},\ \ \ \ \ \xi=4\beta x^2 e^{x N}
\ee
which is the Lambert curve, in accordance with what should be the spectral curve for the Hurwitz theory, \cite{BM,MMkhur}. Since the parameter $x$ provides just a trivial rescaling, we choose it, for the sake of simplicity, equal to 1/2. Hence, the spectral curve finally has the form\footnote{The parameterization of the Lambert curve here is different from that in \cite{BM}, since, in accordance with the definition of the resolvent \cite{AMM}, we choose to work at the vicinity of infinity (at large $z$) with the local parameter ${1\over z}$ instead of $x$ around $x=0$ in \cite[Eq.(2.18)]{BM}, and the leading term is ${1\over z}$. Thus, this formula and \cite[Eq.(2.16)]{BM} are identified upon the redefinition $z\to 1/x$ and $y\to yx$.}
\be\boxed{
y e^{-yz}={\beta e^{N/2}\over z^2}}
\ee

Note that $N$ is now eliminated by the change of variables $p_k\longrightarrow p_k/N$, $z\longrightarrow z/N$,
$\beta \longrightarrow \beta N$.
However, because of additional exponential dependence of $x$ and additional factor of $\beta$ in (\ref{calP0}),
the full $N$-dependence gets very different from that in the $m<0$ models.
In particular, it no longer has any straightforward relation to topological expansion. Because of it, and since the $N$-dependence reduces just to simple rescalings, for the sake of simplicity, we just ignore a possibility of adding $N$-dependent terms in the next subsection.

\subsection{Cut-and-join operators and higher Lambert curves}

Let us note that the ${\cal W}$-operator $\hat {\cal W}_0$, (\ref{W0}) is nothing but the cut-and-join operator $\hat W_{[2]}$ \cite{GJ,MMN}.
Hence, let us consider the next non-trivial generalized cut-and-join operator $\hat W_{[3]}$ \cite{MMN}:
\be
\hat {\cal W}_{[3]}=\sum_{a,b,c\geq 1}^\infty
abcp_{a+b+c} \frac{\p^3}{\p p_a\p p_b\p p_c}
+ \frac{3}{2}\sum_{a+b=c+d} cd\left(1-\delta_{ac}\delta_{bd}\right)
p_ap_b\frac{\p^2}{\p p_c\p p_d} +\nn \\
+ \sum_{a,b,c\geq 1}
(a+b+c)\left(p_ap_bp_c + p_{a+b+c}\right)\frac{\p}{\p p_{a+b+c}}
\ee
As we explained in the previous subsection, we do not need to add any $N$-dependent terms. This operator generates the Hurwitz partition function \cite{MMN}
\be
Z_{[3]}=e^{x\hat {\cal W}_{[3]}}\cdot e^{\beta p_1}=\sum_R\beta^{|R|}S_R\{\delta_{k,1}\}S_R\{p_k\}e^{xC_3(R)}
\ee
where $C_3(R)$ denotes the eigenvalue of second Casimir operator: $C_3(R)=\sum_{R_i}\Big[(R_i-i+1/2)^3-(-i+1/2)^3\Big]$.

In order to construct $\hat{\cal W}^{\rm spec}_{[3]}$, as before, we pick up only the terms with maximal number of derivatives. Then,
\be
\hat{\cal W}^{\rm spec}_{[3]}=\sum_{a,b,c\geq 1}^\infty
abcp_{a+b+c} \frac{\p^3}{\p p_a\p p_b\p p_c}
\ee
and
\be
 e^{{x\over 3}\hat{\cal W}^{\rm spec}_{[3]}} \cdot e^{\beta p_1}=\exp\left(\beta\sum_{n=0}{(2n+1)^{n-1}\over n!}
(x\beta^2)^n {p_{2n+1}\over 2n+1}\right)
\ee
This leads to the spectral curve (we choose $x=1$)
\be
ye^{-z^2y^2}={\beta\over z^2}
\ee
which is the higher Lambert curve.

Similarly, the higher generalized cut-and-join operators $\hat {\cal W}_{[s]}$ \cite{MMN} generate higher Hurwitz partition function \cite{MMN} associated with ``completed cycles" \cite{OP,Lando} (or $r$-spin Hurwitz numbers \cite{Poptr})
\be\label{cc}
Z_{[r]}=e^{x\hat {\cal W}_{[r]}}\cdot e^{\beta p_1}=\sum_R\beta^{|R|}S_R\{\delta_{k,1}\}S_R\{p_k\}e^{xC_r(R)}
\ee
where $C_r(R)$ denotes the eigenvalue of second Casimir operator: $C_r(R)=\sum_{R_i}\Big[(R_i-i+1/2)^r-(-i+1/2)^r\Big]$.
They corresponds to
\be
\hat{\cal W}^{\rm spec}_{[s]}=\sum_{\{a_i\}}\left(\prod_{i=1}^sa_i\right)p_{_{\sum_ia_i}}\frac{\p^s}{\p a_1\ldots\p a_s}
\ee
and
\be
 e^{{x\over k}\hat{\cal W}^{\rm spec}_{[s]}} \cdot e^{\beta p_1}=\exp\left(\beta\sum_{n=0}{((s-1)n+1)^{n-1}\over n!}
(x\beta^{s-1})^n {p_{(s-1)n+1}\over (s-1)n+1}\right)
\ee
which leads to the higher Lambert curve ($x=1$)
\be\boxed{
ye^{-z^{s-1}y^{s-1}}={\beta\over z^2}}
\ee
The higher Lambert curves as the spectral curves for the $r$-spin Hurwitz numbers were earlier discussed in \cite{Poptr}.

Note that one could consider a generic generalized cut-and-join operator $\hat{\cal W}_{\Delta}$ with $\Delta$ being an arbitrary partition \cite{MMN}. However, it acts trivially on $e^{\beta p_1}$ if the partition $\Delta$ has more than one part, or more than one line in terms of Young diagrams. Consider, for instance, $\Delta=[2,1]$. Then,
\be
\hat{\cal W}^{\rm spec}_{[2,1]}=\sum_{a,b=1}ab(a+b-2)p_{a+b}{\p^2\over\p p_a\p p_b}=2p_3{\p^2\over\p p_1\p p_2}+\ldots
\ee
Similarly,
\be
\hat{\cal W}^{\rm spec}_{[2,2]}=\sum_{a,b=1}abc(a+b-2)p_{a+b+c}{\p^3\over\p p_a\p p_b\p p_c}
=2p_4{\p^3\over\p p_1^2\p p_2}+\ldots\nn\\
\hat{\cal W}^{\rm spec}_{[3,1]}=\sum_{a,b=1}abc(a+b+c-3)p_{a+b+c}{\p^3\over\p p_a\p p_b\p p_c}
=2p_4{\p^3\over\p p_1^2\p p_2}+\ldots\nn\\
\hat{\cal W}^{\rm spec}_{[2,1,1]}=\sum_{a,b=1}ab(a+b-2)(a+b-3)p_{a+b}{\p^2\over\p p_a\p p_b}
=6p_4{\p^2\over\p p_1\p p_3}+8p_4{\p^2\over\p p_2^2}+\ldots\nn\\
\hat{\cal W}^{\rm spec}_{[1^k]}=\sum_{a,b=1}a(a-1)...(a-k+1)p_a{\p\over\p p_a}
=k!p_k{\p\over\p p_k}+\ldots
\ee
and, in all these cases,
\be
e^{\hat{\cal W}^{\rm spec}_{\Delta}}  \cdot e^{\beta p_1}= e^{\beta p_1}\ \ \ \ \ \ \Delta\neq [s]
\ee

\subsection{Completed cycles or not?}

One can also consider linear and even multi-linear combinations of $W_\Delta$-operators. Adding lower order operators does not change the $\hat{\cal W}^{\rm spec}$-operator and, hence, does not change the answer for the spectral curve obtained by our procedure. However, among all these combinations, there are some distinguished ones, which provide integrable partition functions \cite{GKM2,MMN,AMMN1}. They are exactly those associated with completed cycles (\ref{cc}). For instance, for $s=1,2,3,4$, these are the operators (one at each level)
\be
\hat W_{[1]},\ \ \ \ \ \hat W_{[2]},\ \ \ \ \ \hat W_{[3]}+{1\over 2}\hat W_{[1]}^2,\ \ \ \ \ \hat W_{[4]}+2\hat W_{[1]}\hat W_{[2]}
\ee
and their arbitrary linear combinations.
We expect that our procedure of getting spectral curves is most immediately applied exactly to such $W$-operators.

Alternatively, one can take {\it other} combinations and claim that, perhaps, integrability is not that necessary for superintegrability of the system, since the partition function ($\alpha_{\Delta,k}$ are arbitrary coefficients)
\be
e^{\sum_\Delta\alpha_{\Delta,k}\hat W_\Delta^k}\cdot e^{\beta p_1}=\sum_R\beta^{|R|}S_R\{\delta_{k,1}\}S_R\{p_k\}
e^{\sum_\Delta\alpha_{\Delta,k}\phi_R(\Delta)^k}
\ee
has a clear superintegrable structure despite not being integrable at generic $\alpha_{\Delta,k}$ \cite{AMMN1}. In this expression,
$\phi_R(\Delta)$ is a peculiarly normalized character of the symmetric group $S_\infty$ \cite{MMN}, and the formula is based on the defining property of the $\hat W_\Delta$-operators \cite{MMN}
\be
\hat W_\Delta S_R=\phi_R(\Delta)S_R
\ee
Thus, these models at $m=0$ and at higher $s$ allows one to study the questions of connections of integrability and superintegrability as well as of relation to the topological recursion and topological expansions, which were difficult to ask in simpler cases. Note that the topological recursion in the completed cycle case was studied earlier in \cite{Poptr}.

\subsection{Spectral densities with $\tr A^k = \delta_{k,s}$}

Note that the higher Lambert curves are surprisingly related to the negative branch of the WLZZ models: their superintegrability relation essentially involves $S_R\{\delta_{k,m}\}$, see (\ref{WLZZnsi}), i.e.,
if the variables $p_k$ are expressed through the matrices $A$, $p_k=\tr A^k$,
it is related to solutions to the equation
\be
\tr A^k = \delta_{k,s}\ \ \ \ \ \hbox{for all}\ k\in \mathbb{Z}_+
\ee
Equivalently, one may ask what are the variables $a_i$, or the eigenvalues of  the matrix $A$ such that
\be
\sum_{i=1}a_i^k = \delta_{k,s}\ \ \ \ \ \hbox{for all}\ k\in \mathbb{Z}_+
\ee
This is a very natural question since the Schur function $S_R$ is a symmetric function just of $a_i$.

In fact, this problem is difficult to solve, however, one may consider the matrix $A$ of a large size $N$, and to study the density of $a_i$ in the large $N$ limit:
\be
\rho(z)dz=\sum_{i=1}^N\delta(z-a_i)
\ee
so that
\be
\int z^k\rho(z)dz=\tr A^k
\ee
Hence, one has to solve the equation
\be
\int z^k\rho(z)dz=\delta_{k,s}
\ee
A solution to this equation is related to a remarkable property at the large $N$ limit \cite{MT}:
the variables $z$ lying on the higher Lambert curve
\be
ze^{-z^s} = w=e^{i\phi}
\label{s-Lambert}
\ee
satisfies the relation
\be
\oint z^{-k} d\phi = \delta_{k,s}
\label{MTrel}
\ee
i.e. only for $k=s$ the series $z^{-k}$ does not have the term $w^0$
(this is non-trivial for all $k=ms$ with $m>1$).

\section{An infinite set of WLZZ models. Positive branch\label{pos}}

\subsection{Description of the models}

The WLZZ proposal at positive integers is to use another pair of operators in order to generate ${\cal W}$-representations, and act with it on an exponential linear in variables $p_k$ instead of unity. More precisely, the procedure is as follows. One starts with the two operators
\be
\hat{\cal W}_1=\sum_{k,l}(k+l+1)p_kp_l{\p\over\p p_{k+l+1}}
+\sum_{k,l}klp_{k+l-1}{\p^2\over\p p_k\p p_l}+
 2N\sum_k(k+1)p_k{\p\over\p p_{k+1}}+N^2{\p\over\p p_1}
\label{Wp1}\\
\hat{\cal W}_2=\sum_{k,l}(k+l+2)p_kp_l{\p\over\p p_{k+l+2}}
+\sum_{k,l}klp_{k+l-2}{\p^2\over\p p_k\p p_l}+
2N\sum_k(k+2)p_k{\p\over\p p_{k+2}}+2N^2{\p\over\p p_2}+N{\p^2\over\p p_1^2}
\label{Wp2}
\ee
which give rise to an infinite set of operators
\be\label{Wpn}
\hat{\cal W}_{m+1}={1\over m}[\hat{\cal W}_m,\hat{\cal W}_1],\ \ \ \ \ m\ge 2
\ee
In fact, these operators can be manifestly described as invariant operators on matrices: with an $N\times N$ matrix $\Lambda$, one can define
\be\label{WL}
\hat{\cal W}_m=\Tr\left({\p^m\over\p\Lambda^m}\right),\ \ \ \ \ m\ge 2
\ee
When acting on invariant functions, i.e. functions of $p_k=\Tr\Lambda^k$, these operators coincide \cite{China} with (\ref{Wpn}).

In fact, these operators can be constructed from the generators of the $\widetilde W$-algebra of sec.\ref{negd}.
Indeed, from relation (\ref{W-tilde-1}) it follows that
\be
\hat{\cal W}_{m}=\Tr\left({\p^{m}\over\p\Lambda^{m}}\right)=\sum_{n=1}p_n\widetilde{W}^{(+,m)}_{n+m}
+N\widetilde{W}^{(+,m)}_{m}
\ee
This implies that the partition function of the corresponding matrix model satisfies the $\widetilde{W}^{(+)}$-constraints (see a particular case in the next subsection). Note that, in variance with the negative branch of WLZZ models, where each degree of $N$ was associated with the $t_0$-derivative, for the positive branch of WLZZ models, $N$ is just $p_0$, and this formula can be rewritten in the form
\be
\hat{\cal W}_{m}=\sum_{n=0}p_n\widetilde{W}^{(+,m)}_{n+m}
\ee
In particular,
\be\label{WI2}
\widetilde{W}^{(+,2)}_n =
\sum_{k=1}(k+n)p_k{\p\over\p p_{k+n}}+\sum_{k=1}^{n-1}k(n-k){\p^2\over\p p_{n-k}\p p_k}
+ Nn\frac{\partial}{\partial p_n},\ \ \ \ \ n\ge 1
\ee

The operators $\hat{\cal W}_m$ (\ref{Wpn}), (\ref{WL}) generate the partition function
\be\label{Zn}
Z_m=e^{\hat{\cal W}_m\over m}\cdot e^{\sum_kg_kp_k/k}
\ee
where $g_k$ are just arbitrary parameters. These partition functions have the superintegrable representations
\be\label{WLZZpsi}
Z_1=\sum_{R,Q}\left({S_R\{N\}S_Q\{\delta_{k,1}\}\over S_Q\{N\}S_R\{\delta_{k,1}\}}\right)^2
S_{R/Q}\{\delta_{k,1}\}S_R\{g_k\}S_Q\{p_k\}\nn\\
Z_{m}=\sum_{R,Q}{S_R\{N\}S_Q\{\delta_{k,1}\}\over S_Q\{N\}S_R\{\delta_{k,1}\}}S_{R/Q}\{\delta_{k,m}\}
S_R\{g_k\}S_Q\{p_k\}\ \ \ \ \ \hbox{at}\ m>1
\ee
where $S_{R/Q}$ is the skew Schur function \cite{Mac}.
This is a KP $\tau$-function in variables $p_k$. It does not come as a surprise, since $e^{\sum_kg_kp_k/k}$ is a KP $\tau$-function, and $\hat W_n$ is an element of $w_\infty$-algebra \cite{MMd,AMMN2}. However, it turns out that this partition function is also a $\tau$-function w.r.t. the second set of variables, $g_k$, which is far less evident. Moreover, even a more strong property is correct: {\bf $Z_n$ is a $\tau$-function of the Toda lattice hierarchy with $N$ being the Toda zero time}. It follows from the fact that it is a KP $\tau$-function to the both sets of time variables\footnote{Note that the traditional choice $t_k$ of time variables of the KP hierarchy as compared with power sums $p_k$ of variables in symmetric functions is $t_k=kp_k$.}  $kp_k$ and $kg_k$, and it satisfies to the lowest Toda-chain hierarchy\footnote{Indeed, one can check that the equation
\be
Z_m(N)\cdot{\p^2 Z_m(N)\over\p p_1\p g_1}-{\p Z_m(N)\over\p p_1}{\p Z_m(N)\over\p g_1}=Z_m(N+1)Z_m(N-1)
\nn
\ee
is satisfied. We checked it up to grading 10 with the computer. This equation along with the KP hierarchies w.r.t. to the two sets of time variables guarantees that $Z_m(N)$ is a $\tau$-function of the Toda lattice hierarchy.}.

Representation (\ref{WLZZpsi}) for the partition function implies that the superintegrability relation for the correlator
\be\label{sip}
\Big<S_Q\{P_k\}\Big>=\sum_{R}{S_R\{N\}S_Q\{\delta_{k,1}\}\over S_Q\{N\}S_R\{\delta_{k,1}\}}S_{R/Q}\{\delta_{k,m}\}
S_R\{g_k\}
\ee
where, as previously, $P_k=\tr M^k$ are traces of matrices at the matrix WLZZ model.

Note that all  the underlined terms in (\ref{Wp1}) and (\ref{Wp2})  break homogeneity in $N$ under the substitution
$p_k\longrightarrow p_k/N$, this is already a signal that they all should be eliminated from $\hat{\cal W}^{\rm spec}$,
see below.

\subsection{\underline{Model with $m=2$}}

Like in the case of negative $m$, for one particular value $m=2$ there is a known one-matrix model realization
\cite{China,China2}.
In this case, this is
the Hermitian matrix model in the external field $\Lambda$ \cite{GKMU}:
\be
Z_2=\int dM \exp\left(-{1\over 2}\Tr M^2+\sum_k {g_k\over k}\cdot\Tr (M+\Lambda)^k\right)
\label{Hmamoext}
\ee
and $p_k=\Tr\Lambda^k$.

Shifting the variable of integration $M\to M-\Lambda$, one can rewrite this partition function in the form
\be\label{W2mamo}
Z_2=e^{-{1\over 2}p_2}\int dM \exp\left(-{1\over 2}\Tr M^2+\Tr M\Lambda +\sum_k {g_k\over k}\cdot\Tr M^k\right)
\ee
When only $g_1$ and $g_2$ are non-vanishing, we have the Gaussian integral,
while when $g_3\neq 0$, we get a more complicated integral, which requires a more advanced approach.

This kind of models was studied in \cite{GKMU}, and this partition function describes a generalized Kontsevich model in the character phase. It satisfies the Ward identities \cite[sec.2.4.2]{GKMU}
\be\label{W2WI}
\left(g_1\delta_{n,1}+\delta_{n,2}-n{\p\over\p p_n}+\sum_{k> 1}g_k\widetilde W^{(+,k-1)}_{k+n-2}\right)\
e^{{1\over 2}p_2}Z_2=0,\ \ \ \ \ n\ge 1
\ee

\subsection{\underline{Spectral curve, $m=2$}}

Let us convert the Ward identities (\ref{W2WI}) with $z^{-s-3}$ and, for the sake of simplicity, preserve only $g_1$, $g_2$ and $g_3$. Then, one gets the spectral curve
\be\label{scW2}
{g_1z^2+g_3Nz^2+g_2z+g_3\over z^6}+{g_3Nz^2+(g_2-1)z+2g_3\over z^3}\Big(y-{N\over z}\Big)
-{g_3N\alpha\over z^3}+g_3\Big(y-{N\over z}\Big)^2=0
\label{scHmamoext}
\ee
where we used (\ref{WI1}) and (\ref{WI2}). Here $\alpha$ is an arbitrary constant, i.e. the spectral curve and the Ward identities have ambiguous solutions (parameterized by one constant).

Indeed, the set of Ward identities (\ref{W2WI}) can be rewritten as
\be
W_n\ Z=\left((g_1+g_3N)\delta_{n,1}+g_2\delta_{n,2}+g_3\delta_{n,3}+(g_2-1)n{\p\over\p p_n}+2g_3(n-1){\p\over\p p_{n-1}}
+g_3N(n+1){\p\over\p p_{n+1}}+\right.\nn\\
\left.+g_3\widetilde{W}^{(+,2)}_n \right)\
Z_2=0,\ \ \ \ \ n\ge 1
\ee
where $\widetilde{W}^{(+,2)}_n$ is given in (\ref{WI2}).

Now, following the general procedure, one converts this infinite set of constraints with powers of $z$, rescales $p_k\to p_k/\hbar$, introduces ${\cal F}:=\hbar^2\log Z$, and rewrites the sum as an equation for the resolvent
\be
\rho(z|p_k):=\hat\nabla_z{\cal F}=\sum_{k\ge 0}{k\over z^{k+1}}{\p {\cal F}\over\p p_k}
\ee
similarly to (\ref{leq}). It remains to puts all $p_k$ zero and take the leading behaviour at small $\hbar$ in order to obtain for $y:=\rho_{-2}(z|0)\Big|_{\hbar\to 0}$  equation (\ref{scW2}).

\subsection{\underline{Spectral curve vs. $\hat{\cal W}^{\rm spec}$, $m=2$}}

Now again let us compare this result with an alternative procedure, which we advocate in this paper.
Namely, keep the highest derivatives terms in (\ref{Wp2}):
\be\label{W2spec}
\hat{\cal W}_2^{\rm spec}=\sum_{k,l}klp_{k+l-2}{\p^2\over\p p_k\p p_l}+
2N\sum_k(k+2)p_k{\p\over\p p_{k+2}}
+\cre{\underline{ 2N^2{\p\over\p p_2}+N{\p^2\over\p p_1^2} }}
\ee
As we shall see, the underlined terms should better be omitted,
i.e. the true definition of $\hat{\cal W}^{\rm spec}$ should be reduced to
terms, which are linear in $p$.
Then,
\be
e^{x\hat{\cal W}_2^{\rm spec}}\cdot e^{g_1p_1}=e^{xNg_1^2}\cdot e^{g_1p_1}
\ee
and
\be
e^{x\hat{\cal W}_2^{\rm spec}}\cdot e^{g_1p_1+\frac{g_2p_2}{2}}=
{\cre\underline{  {e^{xNg_1^2\over 1-2g_2x}\over (1-2g_2x)^{N^2/2}}  }}
\cdot
e^{{1\over 1-2g_2x}\Big(g_1p_1+\frac{g_2p_2}{2}\Big)}
\ee
In fact, in this case, one can just evaluate the Gaussian integral (\ref{W2mamo}) to obtain the r.h.s. of this formula at $x={1\over 2}$. Indeed, the action of $\hat{\cal W}_2^{\rm spec}$ generates the full answer, since the first term in $\hat{\cal W}_2$ (\ref{Wp2}) does not contribute when all $p_k$'s but $p_1$ and $p_2$ are vanishing in the exponential (\ref{Zn}).

The ugly prefactor at the r.h.s., which is independent of times,
is generated by the underlined terms in (\ref{W2spec}).
Omitting them from $\hat{\cal W}_2^{\rm spec}$, we get just
\be\label{g1g2}
e^{x\hat{\cal W}_{2 }^{\rm spec}}\cdot e^{g_1p_1+\frac{g_2p_2}{2}}=
e^{{1\over 1-2g_2x}\Big(g_1p_1+\frac{g_2p_2}{2}\Big)}
\ee
Note that, at the moment, our general principle is to leave in $\hat{\cal W}^{\rm spec}$ only the terms with maximum number of derivatives. However, this principle in all cases considered earlier was equivalent to leaving only terms linear in time variables $p_k$. In the case of $\hat{\cal W}_2^{\rm spec}$ we observe, for the first time, the difference between these two principles, and it becomes clear that we need to follow the second one.

Now note that from (\ref{scW2}) it follows that, in the case of only $g_1$ and $g_2$ non-zero,
\be
y={N\over z}+{g_1z+g_2\over (1-g_2)z^3}
\ee
Inserting this $y$ into $\oint V(xz)\,y(x)  dx$, one gets in the exponential
\be
{\cal P}_2={1\over 1-g_2}\Big(zg_1p_1+\frac{z^2g_2p_2}{2}\Big)
\ee
instead of
\be
{\cal P}'_2={1\over 1-2g_2x}\Big(zg_1p_1+\frac{z^2g_2p_2}{2}\Big)
\ee
in (\ref{g1g2}) after making the substitution $p_k\to z^kp_k$. One could make these two expressions consistent choosing $x={1\over 2}$.

Now consider a more involved case of non-vanishing $g_3$.
To simplify the formulas,
let $g_2=g_1=0$.
Then
\be
\!\!
e^{x\hat{\cal W}_{2}^{\rm spec}}\!\!\cdot e^{g_3p_3/3}=
\exp\!\left(
g_3\sum_{k=1}{(2k)!\over (k+1)!k!}(2g_3x)^{k-1}{p_{k+2}\over {k+2}}
+ N \sum_{k=1} {(2k-1)!\over k!(k-1)!}(2g_3x)^k{p_k\over k}
+\right.\nn\\
+ \left.\underline{\sum_{m=1}(Ng_3^2x^3)^m\cdot  \sum_{k=1}N\alpha^{(m)}_k \cdot (2g_3x)^k \frac{p_k}{k}}   \right)
\label{expW2}
\ee
Numeric coefficients $\alpha^{(m)}_k$ are quite complicated,
but all the underlined terms come with extra powers of $Ng_3^2$.
This provides a selection rule, which allows one to eliminate them in a regular way.

Now let us again choose $x={1\over 2}$. Then, the spectral curve is associated with the main terms, which are not underlined is (in accordance with ${\cal P}_2=\oint V(xz)\,y(x)  dx$)
\be\label{scfW2}
y_2(z) =\sum_{k=1}{2k!\over (k+1)!k!}{g_3^k\over z^{k+3}}+N \sum_{k=1} {(2k-1)!\over k!(k-1)!}{g_3^k\over z^{k+1}}=\nn\\
={1\over 2g_3z^2}\left(1-\sqrt{1-\frac{4g_3}{z}}\right)-{1\over z^3}
+{N\over 2z}\left(\frac{1}{\sqrt{1-\frac{4g_3}{z}   }}+1\right)+O(N^2)
\ee
In order to compare this curve with (\ref{scW2}) note that, like in s.\ref{00model}, elimination of  the underlined terms in (\ref{W2spec}) is done by the rescaling $p_k\longrightarrow p_k/N$ and considering small $N$ limit. Hence, the relation of topological and $1/N$-expansion breaks down (similarly to what happened in s.\ref{00model}).

Consider now the leading order of (\ref{scW2}) at {\bf small} $N$:
\be\label{sclo}
{g_3\over z^6}+{2g_3-z\over z^3}y^{(0)}+g_3\left(y^{(0)}\right)^2=0
\ee
Its solution is exactly the curve (\ref{scfW2}) at small $N$, the first two terms:
\be
y^{(0)}(z) ={1\over 2g_3z^2}\left(1-\sqrt{1-\frac{4g_3}{z}}\right)-{1\over z^3}=y_2^{(0)}(z)
\ee
Now, one can consider the first small $N$ correction to (\ref{scW2}). It gives rise to a more complicated formula than just
\be\label{y1c}
y_2^{(1)}(z)={N\over 2z}\left(\frac{1}{\sqrt{1-\frac{4g_3}{z}   }}+1\right)
\ee
in (\ref{scfW2}):
\be\label{y1}
y^{(1)}(z)={N\over 2z}\left(\frac{1-2\alpha g_3+{2g_3\over z}(\alpha-1)}{\sqrt{1-\frac{4g_3}{z}   }}+1\right)
\ee
However, note that the rescaling $p_k\longrightarrow p_k/N$ would imply
also the rescaling $g_k\longrightarrow Ng_k$ in order to preserve exponential intact. This means that one also has to consider a leading behaviour at small $g_3$. The leading contribution at small $g_3$ in (\ref{y1}) (after the rescaling of $z\to g_3z$) is exactly (\ref{y1c}) upon the choice of $\alpha=1$: $y^{(1)}(z)\to y_2^{(1)}(z)$, and finally we come to (\ref{scfW2}).

Note that one can introduce new variables $Y_2=z^3y_2^{(0)}$ and $x=Z/g_3$ such that the spectral curve (\ref{sclo}) becomes
\be\boxed{
(ZY_2)^{1/2}-Y_2-1=0}
\ee

To summarize, we see that the spectral curve (\ref{scHmamoext}),
which can be extracted from the matrix model realization (\ref{Hmamoext}),
is consistent with our universal definition from
the $p$-linear part  $\hat W_{\rm spec}$ of the $\hat W$ operator. However,
in this case, one needs to deal with the small $N$ limit, and the idea of large $N$ expansion,
which continued to be safe for the WLZZ
models with $m<0$ needs to be changed for its opposite at $m>0$.

\bigskip

\begin{center}
\framebox{\parbox{12cm}{\bf Thus, we finally can formulate the general prescription: in order to construct the operator $\hat {\cal W}^{\rm spec}$, one has to leave in the original operator $\hat W$ linear in $p_k$ terms with maximum number of derivatives.}}
\end{center}

\bigskip

As for the large or small $N$ limit and the topological expansion, as we demonstrated, it depends on the concrete model.

\subsection{Spectral curve in the $m=1$ model}

Now consider models at other values of $m$. We consider only the leading small $N$ order of the spectral curve.

We start with the very first example, $m=1$. In this case, the $\hat W$-operator is given by formula (\ref{Wp1}), and, in accordance with our general rule,
\be
\hat{\cal W}_1^{\rm spec}=\sum_{k,l}klp_{k+l-1}{\p^2\over\p p_k\p p_l}+
\cb{ \underline{N^2{\p\over\p p_1}}}
\ee
Since we deal with the small $N$ limit, we drop out the underline term. Then,
\be
e^{x\hat {\cal W}_1^{\rm spec}}\cdot e^{g_1p_1+g_2p_2/2}=\exp\left({g_1\over 1-xg_1}p_1+
{1\over x(1-xg_1)}\sum_{n\ge 1}{1\over n}C^{3n}_{n-1}\left({xg_2\over  (1-xg_1)^{3}}\right)^n{p_{n+1}\over n+1}\right)
\ee
where $C^n_k$ are the binomial coefficients.

In order to get the spectral curve, we choose $x=1$. Then, the spectral curve is associated with the leading term at small $N$ (in accordance with ${\cal P}_1=\oint V(xz)\,y(x)  dx$)
\be\label{g1}
y^{(0)}_1={1\over (1-g_1)z^2}\left(g_1+\underbrace{
\sum_{n\ge 1}{1\over n}C^{3n}_{n-1}\left({g_2\over z(1-g_1)^3}\right)^n}_{Y_1-1}\right):={1\over (1-g_1)z^2}(g_1-1+Y_1)
\ee
Upon introducing also a new variable $Z=z(1-g_1)^3/g_2$, this sum satisfies the equation for the spectral curve
\be
Z^{-1}Y_1^3-Y_1-1=0
\ee
Thus, one can see that the role of $g_1$ is basically to rescale $g_2$ for $g_2/(1-g_1)^3$, much similar to the rescaling with $1/(1-g_2)$ in the $m=2$ case.

Hence, now we drop $g_1$, and switch on the $g_3$ parameter instead:
\be\label{g2g3}
e^{x\hat {\cal W}_1^{\rm spec}}\cdot e^{g_2p_2/2+g_3p_3/3}=\exp\left({1\over x}\sum_{n\ge 1,k\ge 0}
{1\over n}C^{3n+2k}_{n-1}C^{n}_{k}(xg_2)^{n-k}(xg_3)^k{p_{n+k+1}\over n+k+1}\right)\nn\\
\ee

The equation for the spectral curve for
\be
y_1^{(0)}=\sum_{n\ge 1,k\ge 0}{1\over n}C^{3n+2k}_{n-1}C^{n}_{k}g_2^{n-k}g_3^kz^{-n-k-2}={Y_1-1\over z^2}
\ee
is rather simple:
\be
{g_3\over z^2}Y_1^5+{g_2\over z}Y_1^3-Y_1+1=0
\ee
One can also easily restore the parameter $g_1$ in (\ref{g2g3}):
$$
e^{x\hat {\cal W}_1^{\rm spec}}\cdot e^{g_1p_1+g_2p_2/2+g_3p_3/3}=
$$
\be
=\exp\left({g_1\over 1-xg_1}p_1+
{1\over x(1-xg_1)}\sum_{n\ge 1,k\ge 0}
{1\over n}C^{3n+2k}_{n-1}C^{n}_{k}\left({xg_2\over (1-xg_1)^{3}}\right)^{n-k}
\left({xg_3\over (1-xg_1)^{5}}\right)^k{p_{n+k+1}\over n+k+1}\right)
\ee
It again reduces to the rescalings $g_2\to g_2/(1-xg_1)^3$, $g_3\to g_3/(1-xg_1)^5$.

Now the general formula is clear: adding on more parameters $g_k$, $k=2,\ldots,K$ gives rise to the spectral curve
\be\boxed{
\sum_{k=2}^K{g_k\over z^{k-1}}Y_1^{2k-1}-Y_1+1=0}
\ee
and, the parameters rescalings upon switching on $g_1$ are: $g_k\to g_k/(1-g_1)^{2k-1}$, $k=1,\ldots,K$.

\subsection{Spectral curve in the $m=3$ model}

Our next example is $m=3$, and the $\hat{\cal W}_3$-operator is
\be
\hat{\cal W}_3=  \sum_{a,b,c\ge 1}abcp_{a+b+c-3}\frac{\p^3}{\p p_a\p p_b\p p_c}+
\sum_{b,c\ge 1}\sum_{a=1}^{b+c+2} a(b+c-a+3)p_bp_c{\p^2\over\p p_a\p p_{b+c-a+3}}+\nn\\
+\sum_{b,c\ge 1}\sum_{a=1}^{b+1} a(b+c-a+3)p_bp_c{\p^2\over\p p_a\p p_{b+c-a+3}}+
\sum_{a,b,c\ge 1}(a+b+c+3)p_ap_bp_c{\p\over \p p_{a+b+c+3}}+\nn\\
+3N\sum_{b\ge 1}\sum_{a=1}^{b+2} a(s
b-a+3)p_b{\p^2\over\p p_a\p p_{b-a+3}}
+3N\sum_{a,b\ge 1} (a+b+3)p_ap_b{\p\over\p p_{a+b+3}}\nn\\
+ \sum_{a\ge 1}{(a+1)(a+2)(a+3)\over 2}p_a{\p\over\p p_{a+3}}
+3N^2\sum_{a\ge 1}(a+3)p_a{\p\over\p p_{a+3}}+\nn\\
+N{\p^3\over\p p_1^3}+6N^2{\p^2\over\p p_1\p p_2}+3N(N^2+1){\p\over\p p_3}\nn\\
\ee
To get the operator $\hat{\cal W}_3^{\rm spec}$, in accordance with the general principle, we leave only the first term in this expression:
\be
\hat{\cal W}_3^{\rm spec} =\sum_{a,b,c}abcp_{a+b+c-3}\frac{\p^3}{\p p_a\p p_b\p p_c}
\label{W3spec}
\ee
In this case, the action of $e^{x\hat{\cal W}_3^{\rm spec}}$ on $e^{g_1p_1}$ is trivial:
\be\label{sc31}
e^{x\hat{\cal W}_3^{\rm spec}}\cdot e^{g_1p_1}=e^{g_1p_1}
\ee
while the actions of $e^{x\hat{\cal W}_3^{\rm spec}}$ on $e^{\frac{g_2p_2}{2}}$ and $e^{\frac{g_3p_3}{3}}$ are
\be\label{sc32}
e^{x\hat{\cal W}_3^{\rm spec}}\cdot e^{\frac{g_2p_2}{2}} =
\exp\left(\sum_{n=0}{g_2\over n+1}C^{2n}_{n}(xg_2^2)^{n}{p_{n+2}\over n+2}\right)\\
e^{x\hat{\cal W}_3^{\rm spec}}\cdot e^{\frac{g_3p_3}{3}} =
\exp\left(\sum_{n=0}{2g_3\over 3n+2}C^{4n+1}_{n}(xg_3^2)^{n}{p_{3n+3}\over 3n+3}\right)
\label{sc33}
\ee
Switching on the $g_1$ parameter in these cases, as previously, just rescales the parameters $g_k$. For instance, $g_2\to g_2/\sqrt{1-4xg_1g_2}$. This also adds a contribution proportional to $p_1$ similar to $\displaystyle{g_1\over (1-g_1)z^2}$ in (\ref{g1}).

The spectral curves associated with (\ref{sc32}) and (\ref{sc33}) are accordingly
\be
{z^3\over g_2}y_3=1+z^5y_3^2\nn\\
{z^4\over g_3}y_3=(1+z^5y_3^2)^2
\ee
where we omitted the superscript $0$ of $y^{(0)}$, for the sake of brevity.

Now the natural conjecture is that, for a non-zero parameter $g_k$, the curve looks like
\be\label{sc3g}
{z^{k+1}\over g_k}y_3=(1+z^5y_3^2)^{k-1}
\ee
Indeed, let us consider the action of $e^{x\hat{\cal W}_3^{\rm spec}}$ on $e^{\frac{g_4p_4}{4}}$:
\be
e^{x\hat{\cal W}_3^{\rm spec}}\cdot e^{\frac{g_4p_4}{4}}=
\exp\left\{\sum_{k=1}{3(6k-4)!\over (k-1)!(5k-2)!}(3x)^{k-1}g_4^{2k-1}{p_{5k-1}\over 5k-1}\right\}
\ee

In order to get the spectral curve, we choose $x={1\over 3}$. Then, the spectral curve is associated with the leading term at small $N$ is (in accordance with ${\cal P}_3=\oint V(xz)\,y(x)  dx$)
\be\label{scW3}
y_3={3\over g_4}\sum_{k=1} {(6k-4)!\over (k-1)!(5k-2)!}\left({g_4^2\over z^5}\right)^k
\ee
This sum satisfies the equation for the spectral curve (\ref{sc3g}).

At last, when the first $K$ parameters $g_k$ are non-zero, the spectral curve is
\be\label{sc3gm}\boxed{
y_3=\sum_{k=2}^K{g_k\over z^{k+1}}(1+z^5y_3^2)^{k-1}}
\ee

Note that, upon introducing new variables $Y_3=g_4y^{(0)}_3$, $Z=z^5/g_4^2$, sum (\ref{scW3}) can be rewritten in the form
\be\boxed{
(ZY_3)^{1/3}-ZY_3^2-1 = 0}
\ee

\subsection{Generic $m$}

Generalization to all the WLZZ models with arbitrary $m>0$ is now straightforward.
For any $m>0$, relevant in  $\hat{\cal W}_m^{\rm spec}$ is just the term:
\be
\hat {\cal W}_m^{\rm spec} =\sum_{\{a_i\}}\left(\prod_{i=1}^ma_i\right)p_{_{\sum_ia_i-m}}\frac{\p^m}{\p a_1\ldots\p a_m}
\ee
In the most interesting case, one gets
\be
e^{x\hat{\cal W}_m^{\rm spec}}\cdot e^{\frac{g_{m+1}p_{m+1}}{m+1}}
= \exp\left\{\sum_{k=0}{1\over n_k}C^{n_k}_{k}m(mx)^kg_{m+1}^{(m-1)k+1}\cdot{p_{n_k-k+1}\over n_k-k+1}
\right\}
\ee
where $n_k=(m-1)(mk+1)+1$.

As before, in order to get the spectral curve, we choose $x={1\over m}$. Then, the spectral curve is associated with the leading term at small $N$ (in accordance with ${\cal P}_m=\oint V(xz)\,y(x)  dx$)
\be
y_m=g_{m+1}^{2-m}z^{(m+1)(m-3)}
\sum_{k=0}{m\over n_k}C^{n_k}_{k}\left({g_{m+1}^{m-1}\over z^{m^2-m-1}}\right)^{k+1}
\ee
Upon introducing new variables $Y_m=g_{m+1}^{m-2}z^{(m+1)(3-m)}y^{(0)}_m$, $Z=z^{m^2-m-1}/g_{m+1}^{m-1}$, this sum satisfies the equation for the spectral curve
\be\boxed{
(ZY_m)^{1/m}-Z^{m-2}Y^{m-1}-1 = 0}
\ee
At the same time, the counterpart of (\ref{sc3gm}) is
\be\label{scmg}\boxed{
y=\sum_{k=2}^K{g_k\over z^{k+1}}(1+z^{2m-1}y^{m-1})^{k-1}}
\ee

\section{Exponentiation principle\label{CH}}

We did not comment so far a miraculously looking property that action of the $\hat {\cal W}^{\rm spec}$ operator on the exponential linear in $p_k$'s produces also an exponential linear in $p_k$'s. In fact, this property follows from the Campbell-Hausdorff formula (CHF) as we will discuss now.

Consider first the simplest case. Even getting the formula
\be
\exp\left(\sum_{k=1}^\infty kp_{k+1}\frac{\p}{\p p_k} + Np_1\right)\cdot 1
= \exp\left(N\sum_{k=1}\frac{p_k}{k}\right)
\label{eqn1}
\ee
 requires a few steps.

In fact, it results from a multiple application of the CHF, e.g.
\be
\exp\left(y\frac{\p}{\p x} + Nx\right)\cdot 1 =
e^{-\frac{Ny}{2}} e^{y\frac{\p}{\p x}} e^{Nx}\cdot 1
= e^{-\frac{Ny}{2}} e^{N(x+y)} = e^{\frac{Ny}{2} + Nx}
\ee
where we used the CHF in the form
\be
e^{A+B} = e^{A}e^B e^{-\frac{[A,B]}{2}}
\ee
which is valid when the commutator $[A,B]$ commutes with both $A$ and $B$.
Further,
\be
\exp\left(2z\frac{\p}{\p y}+ y\frac{\p}{\p x} + Nx\right)\cdot 1 =
e^{2z\frac{\p}{\p y}} e^{y\frac{\p}{\p x}} e^{-z\frac{\p}{\p x}}e^{-\frac{Ny}{2}}e^{\frac{Nz}{3}}  e^{Nx}\cdot 1
=  e^{2z\frac{\p}{\p y}} e^{-\frac{2Nz}{3}+\frac{Ny}{2}+Nx}
= e^{\frac{Nz}{3}+\frac{Ny}{2}+Nx}
\ee
and so on. At this stage, we used the CHF in the form
\be
e^{A+B+C} = e^{A}e^B e^{-\frac{[A,B]}{2}}e^{-\frac{[B,C]}{2}}e^{\frac{[[A,B],C]}{3}}e^C
\ee
where $[[A,B],C]$ commutes with all other quantities in this formula.

One can easily change the weights in (\ref{eqn1}):
\be
\exp\left(\sum_{k=1}^\infty c_k p_{k+1}\frac{\p}{\p p_k} + Np_1\right)\cdot 1
= \exp\left(N\sum_{k=1}\frac{ c_1\ldots c_{k-1}p_k}{k!}\right)
\label{eqn1c}
\ee
However, if one attempts to substitute the
exponential functions by anything else:
\be
G\left(\sum_{k=1}^\infty kp_{k+1}\frac{\p}{\p p_k} + Np_1\right)\cdot 1
= H\left(N\sum_{k=1}\frac{p_k}{k}\right)
\ee
there will be no solutions different from $H(x)=G(x)=e^x$.
In this sense, {\bf the exponential function is distinguished}.

Our example demonstrates that, since $\hat{\cal W}^{\rm spec}$ is linear in $p_k$'s though may involve higher derivatives, $\exp\left(\hat{\cal W}^{\rm spec}\right)$ upon acting on unity produces an exponential linear in $p_k$'s. In more involved examples of $\exp\left(\hat{\cal W}^{\rm spec}\right)$, the calculations are more tedious, however, they work same way. An even more complicated case is when $\exp\left(\hat{\cal W}^{\rm spec}\right)$ is acting on $\exp\left(\sum_kg_kp_k/k\right)$. However, the result is still an exponential linear in $p_k$'s. In order to prove this, one has to use the Dynkin form of the CHF \cite{Dn,Dnb}:
\be\label{CHF}
\exp(\hat A)\cdot\exp(\hat B)=\exp\left(\sum_n{(-1)^n\over n}\sum_{\{r_i+s_i>0\}}{1\over\prod_{i=1}^n
r_i!s_i!\cdot\sum_{i=1}^n(r_i+s_i)}\times\right.\nn\\
\left.\times\underbrace{[\hat A,[\hat A,\ldots[\hat A,}_{r_1}\underbrace{[\hat B,[\hat B,\ldots[\hat B,}_{s_1}\ldots
\underbrace{[\hat A,[\hat A,\ldots[\hat A,}_{r_n}\underbrace{[\hat B,[\hat B,\ldots ,\hat B]}_{s_n}]]\ldots ]\right)
\ee
where $[\hat X]:=\hat X$.

It is clear from this formula that $\exp\left(\hat{\cal W}^{\rm spec}\right)\cdot \exp\left(\sum_kg_kp_k/k\right)$
contains only commutators of operators of the form $\sum p_k\hat D_k$, where $\hat D_k$ is a pure differential operator of a finite order, and commutators of these operators have also this form. Hence, we ultimately come to conclusion that
$\exp\left(\hat{\cal W}^{\rm spec}\right)\cdot \exp\left(\sum_kg_kp_k/k\right)=\exp\left(\sum p_k\hat D_k\right)\cdot 1$, and we return to our example above.

\section{Comments}

In this subsection, we mention some other promising directions for further development of the spectral curve theory for the WLZZ models. We do not elaborate on any of them, but hopefully they will attract attention in the future.

\subsection{${\cal W}$-representation vs $W$-constrains}

Actually, ${\cal W}$-representation (\ref{wrep}) is naturally exponential,
if $Z\{p\}$ satisfies the $W$-constraints \cite{Max,MMMR,MMM}.

For instance, as we discussed in secs.2-3 (Eqs.(\ref{W2}),(\ref{sum1}), etc), the ${\cal W}$-representation in the negative branch WLZZ models is of the form
\be
\hat{\cal W}_{-m} = \sum_{k=1}^\infty  p_k \widetilde {W}_{k-m}^{(-,m)}
\ee
This allows one immediately to obtain the ${\cal W}$-representation: since the partition function $Z_{-m}$ satisfies the set of $\widetilde {W}^{(-,m)}$-algebra constraints (\ref{Wmcon}),
\be\label{Wcon}
\left(k\frac{\p}{\p p_{k}} -\widetilde {W}_{k-m}^{(-,m)}\right) Z = 0
\ee
summing it up with $p_k$, we obtain
\be\label{Weq}
\Big(\hat l_0 - \underbrace{\sum_k p_k \widetilde {W}_{k-m}^{(-,m)}\{p\}}_{\hat {\cal W}_{-m}\{p\}}\Big)Z\{p\}=0
\ee
with the grading operator $\hat l_0 := \sum_k kp_k\frac{\p}{\p p_k}$,
and with $\hat {\cal W}_{-m}$ having a given grading, $m$ so that
\be
[\hat l_0,\hat{\cal W}_{-m}]=m\hat{\cal W}_{-m}
\ee
Now, it is immediate to prove that
\be
Z= e^{{1\over m}\hat{\cal W}_{-m}} \cdot 1
\ee
satisfies (\ref{Weq}). Since the solution to (\ref{Wcon}) is unique, which can be checked following the line of \cite{AMM}, (or, equivalently, the solution to (\ref{Weq}) is unique, which can be checked following the line of \cite{Max1}), we obtain that (\ref{Wcon}) are equivalent to the $W$-representation.

This is a generalization of the elementary fact
\be
\left( x\frac{d}{dx} - x^m\right) Z = 0  \ \ \ \Longrightarrow \ \ \
Z \sim e^{x^m/m}
\label{eqxs}
\ee

An interesting question is if we know a single $\hat{\cal W}$ in a more generic situation,
can we find the entire set of constraints (\ref{Wcon})? In particular, what is the set of constraints in
the positive branch WLZZ models?
This question stands from \cite{MMkhur}, where it was shown that
a very simple $\hat{\cal W}$ is associated with somewhat non-trivial,
``conjugate or deformed continuous" Virasoro constraints.

\subsection{Large $N$ limit of superintegrability relation}

In this subsection, we again return to the superintegrability relation (\ref{sin}) in the case of  the negative branch of the WLZZ models.

For the sake of definiteness, we start with the Gaussian case $m=2$. In this case, the resolvent (\ref{rhoGauss}) satisfies the spectral curve equation (\ref{sc2}), and its imaginary part (jump at the branch cut) $\rho(z)={1\over 2\pi i}\lim_{\epsilon\to 0}\Big(y(z-i\epsilon)-y(z+i\epsilon)\Big)={\Im y(z)\over 2\pi}\sim\sqrt{4N-z^2}$ is sometimes called
spectral density since it provides the distribution of eigenvalues \cite{Mehta,TE,ChMMV} reasonable at large $N$, when multi-trace correlators factorize. This means that
\be
\Big<P_{k_1}P_{k_2}\Big>_\infty =\Big<P_{k_1}\Big>_\infty \Big<P_{k_2}\Big>_\infty
\ee
and
\be
\Big<P_k\Big>_\infty = \int z^k\rho(z)dz
\ee
It is instructive to see how the superintegrability relations (\ref{sin}) trivialize in this limit.

Since in Gaussian case $\Big<P_{2k}\Big>_\infty\sim N^{k+1}$,
dominating in the Schur average is the item with maximal number of $P_2$:
\be
\Big<S_R\Big>_\infty = \Big<S_R\{\delta_{k,2}\}\cdot P_2^{|R|/2}\Big>_\infty =
S_R\{\delta_{k,2}\}\cdot \Big<P_2\Big>_\infty^{|R|/2}=N^{|R|} S_R\{\delta_{k,2}\}
\ee
At the same time, this is exactly the large $N$ limit of
the r.h.s. of the superintegrability relation:
\be
S_R\{\delta_{k,2}\} \frac{S_R\{N\}}{S_R\{\delta_{k,1}\}}
\ {\stackrel{N\to\infty}{\longrightarrow}}\ N^{|R|}\  S_R\{\delta_{k,2}\}
\ee
since dominating is the contribution from the maximal power of $p_k=N$,
which is $p_1^{|R|}$.

The main point is that the superintegrability relation in the large $N$ limit is just trivial:
no requirements are imposed on actual values of $\Big<P_{2k}\Big>_\infty$ for other $k\neq 1$.
Completely the same consideration can be repeated for any negative branch WLZZ model.

To put it differently, the superintegrability relations in the large $N$ limit become
linear in the sector with definite grading $|R|$:
\be
\Big< S_R \Big>_\infty = N^{|R|} S_R\{\delta_{k,m}\}
\ee
This means that they are not longer restricted to characters, one can take any linear
combination of $S_R$ with the same $|R|$:
\be
\Big< F \Big>_\infty = N^{|R|}F\{\delta_{k,m}\}
\ee
for any $ F = \sum_{R \ {\rm with\ a\ given}\ |R|}  f_RS_R$.
In particular, one obtains a factorization:
since
\be
S_{R_1}S_{R_2} = \sum_{R_3:\ |R_3|=|R_1|+|R_2|} {\cal N}^{R_3}_{R_1R_2} S_{R_3}
\ee
one gets
\be
\Big<S_{R_1}S_{R_2}\Big>_\infty = \sum_{R_3}  {\cal N}^{R_3}_{R_1R_2} \Big<S_{R_3}\Big>_\infty
= N^{|R_1|+|R_2|} \sum_{R_3}  {\cal N}^{R_3}_{R_1R_2} S_{R_3}\{\delta_{k,m}\} = \nn \\
= N^{|R_1|+|R_2|} \cdot S_{R_1}\{\delta_{k,m}\} S_{R_2}\{\delta_{k,m}\}
= \Big<S_{R_1} \Big>_\infty\Big< S_{R_2}\Big>_\infty
\ee

\subsection{Large $N$ limit of double averages}

Note that the factorization of correlators at large $N$ should not be taken for granted. Consider, for instance, double averages in the negative branch WLZZ models that are factorized due to the superintegrability \cite{MMd,MMNek}. These correlators are generated by the action of the $W$-operators $\hat {\cal W}_m$ on the Schur function $S_R$ as functions of $P_k$,
\be\label{da}
\Big<S_Q\{\hat {\cal W}_k\}\cdot S_R\{P_k\}\Big>_{WLZZ_{-m}}=
{\displaystyle{S_{R/Q}\{\delta_{k,m}\}S_R\{N\}}\over \displaystyle{S_R\{\delta_{k,1}\}}}
\ee
It is curious that, though these $W$-operators generate the positive branch of the WLZZ models, the correlators we are talking about are those in the negative branch models.

As we demonstrated in \cite{MMd} for the Gaussian ($m=2$) model, the averages (\ref{da}) can be reduced to a correlator of the form
\be
\Big<S_Q\{\hat {\cal W}_k\}\cdot S_R\{P_k\}\Big>=\Big<K_Q\{P_k\}\cdot S_R\{P_k\}\Big>
\ee
where the polynomials $K_R$ form a complete basis, and celebrate the property
\be
\Big<K_R\cdot K_Q\Big>=\displaystyle{S_{R}\{N\}\over \displaystyle{S_R\{\delta_{k,1}\}}}\delta_{RQ}
\ee
Examples of these polynomials can be found in \cite[Appendix]{MMd}\footnote{Note that, throughout the paper \cite{MMd}, we discussed another basis of polynomials, $K_\Delta$, the two related by the Fr\"obenius formula
\be
K_R=\sum_\Delta{\psi_R(\Delta)\over z_\Delta}K_\Delta\nn
\ee
where $\psi_R(\Delta)$ is the symmetric group character, and $z_\Delta$ is the standard symmetric factor of the Young diagram (order of the automorphism) \cite{F}.
}.

Now the point is that these double averages are not factorized. This is because the operators
do not have a definite grading.
Moreover, terms of different gradings come with $N$-dependent coefficients
and are carefully matched to cancel the $N$-dependent contributions.
In result, the average
$\Big<K_Q S_{R}\Big> $
does not grow as $N^{|Q|+|R|}$, it is rather $\sim N^{|R|}$.
Moreover,  $\Big<K_Q P_{R}\Big> $  for individual time-variables $P_R$ can grow even slower,
e.g.  $\Big<K_{[1,1]} P_{2}\Big> =0 $,
while  $\Big<  P_{2}\Big>  = N^2$.

This is consistent with the fact that the ``eigenvalues" $\mu$ in
\be
\Big<K_Q S_{R}\Big> = \mu_{Q,R}\cdot\Big<  S_{R}\Big>,\ \ \ \ \ \mu_{Q,R}={\displaystyle{S_{R/Q}\{\delta_{k,2}\}\over
S_{R}\{\delta_{k,2}\}}}
\ee
do not depend  on $N$ (instead of growing like $N^{|Q|}$).

\section{Conclusion}

The main goal of this paper was to learn how the spectral curve
for the resolvent $y(z)$
emerges from the ${\cal W}$-representation of the partition function.
We demonstrated that, in the standard examples of matrix models,
it is described by a truncated version of ${\cal W}$-operator,
$\hat{\cal W}^{\rm spec}$. In order to construct the operator $\hat {\cal W}^{\rm spec}$ from the full $\hat{\cal W}$, one has to leave in $\hat{\cal W}$ linear in $p_k$ terms with maximum number of derivatives (taking into account that the coefficient $N$ is also a derivative with respect to a variable $t_0$). Then,
\be
e^{\hat {\cal W}^{\rm spec}}\cdot 1 = \exp\left({\cal P}\right)
\label{Wrepconc}
\ee
and ${\cal P}$ is {\bf linear} in time variables.
We explained in sec.\ref{CH} why this linearization happens exactly to the {\bf exponential} functions
on the both sides of (\ref{Wrepconc}).
Now, the substitution $p_k\to z^kp_k$ makes ${\cal P}$ depending on a spectral parameter $z$, and allows one to generate the function (resolvent) $y(z)$ such that
\be
{\cal P}(z)=\oint V(xz)\,y(x)  dx
\ee
where $V(z)=\sum_kp_kz^k/k$ is the matrix model potential. The resolvent $y(z)$ satisfies the spectral curve equation.

As a highly non-trivial check of this conjecture, we applied it
to the intriguing family
of WLZZ models \cite{China}, which so far were defined only through
${\cal W}$-representations.
These models have a parameter $m$, which characterizes at once
the grading of the operator and the maximal number of derivatives,
the two {\it a priori} independent parameters.
Somehow their identification seems to provide an especially interesting
class of partition functions,
which possess, apart from integrability, also a simple superintegrability property.
We showed that, for negative $m$, the above prescription for
$\hat{\cal W}^{\rm spec}$ is just the correct one and leads
to the family of spectral curves
\be
y^{|m|} - zy+ N = 0
\ee
which generalize the one for the case of $m=-2$, the ordinary Hermitian model at the Gaussian point.

In order to check it, we needed to restore the $W$-constraints on the partition function in this case:
\be
\left(n\frac{\p}{\p p_{n}} +\widetilde {W}_{n-m}^{(-,m)}\right) Z = 0
\ee
where the operators $\hat W_n^{(m)}$ are obtained from the relation
\be
\hat{\cal W}_{-m} = \sum_{k=1}^\infty  p_k \widetilde {W}_{k-m}^{(-,m)}
\ee
and then to construct the loop equations, the leading large $N$ behaviour of them just giving rise to the spectral curve.

However, for positive $m$ the situation is more complicated: in this case, there is no large $N$ topological expansion, and the spectral curve limit rather corresponds to the small $N$ limit. On the other hand, there is neither known a set of the $W$-constraints on the partition function. Hence, though, by our general procedure, we obtained the spectral curve
\be
y=\sum_{k=2}{g_k\over z^{k+1}}\Big(1+z^{2m-1}y^{m-1}\Big)^{k-1}
\ee
we could check that it coincides with the correct one only in the case of $m=2$ when there exists a realization of the partition function as the Gaussian matrix model in the external field \cite{China,GKMU}.

At the boundary between positive and negative $m$ lies the case of $m=0$,
where it still makes sense to untie the number of derivatives $s$
from the grading $m=0$.
This gives rise to a whole family of Lambert spectral curves
\be
ye^{-z^{s-1}y^{s-1}}={\beta\over z^2}
\ee
of which $s=1$ is the standard example of the Hurwitz model \cite{BM,MMkhur}.
In this case, the $N$ dependence is already not quite simple, and
the spectral curve is not described by a naive large $N$ limit
(one should rather substitute $\beta\to e^{-N/2}\beta$).

To conclude, our approach allows one to construct the spectral curves for the WLZZ models.
The point is that the model defined via a ${\cal W}$-representation may be nice
(in particular, superintegrable), and a nice expression is available
for the would-be spectral curve even if a matrix model representation, or even a set of $W$-constraints on the partition
function are unavailable.
Moreover, the spectral curves are not obligatory related to the large $N$ limit:
the positive branch of WLZZ models is rather associated with the {\bf small} $N$ limit.
What this means for the topological expansion and topological recursion still remains
to be understood.

This study provides new insights into the notion of spectral curve,
and thus of the AMM-EO topological recursion \cite{AMM,AMMtr,EO}.
It is an interesting question how the later one is constructed from ${\cal W}$-representations,
and what are the restrictions (if any) on the possible choice of $\hat{\cal W}$
and the ``vacuum" state.
This is also related to the ambiguity problem of ${\cal W}$-representations \cite{AMM,Max1}.

One of the straightforward generalizations of this investigation can be
to confirm our general recipe for generating the spectral curve in the case of $\beta$-deformations,
which are readily available for the WLZZ models \cite{China}.

To summarize, the WLZZ models provide us with entire three {\it families} of superintegrable theories: for $m<0;$ $m=0,\ s\ge 2$ and $m>0$, which generalize known and rather non-trivial examples at $m=\pm 2$ and $m=0,\ s=2$.  This opens an opportunity of studying problems that could not be fully addressed before, like relation between super- and ordinary integrability (seemingly broken for $s>2$), or relation between the spectral curves and the topological recursion and the large $N$ expansion (broken at $m>0$), or relation between the $W$-representations and the Ward identities. This makes further study of these models very promising and challenging.
At the same time, it remains unclear what makes these models so successful, and which properties of the ${\cal W}$-operator are responsible for (super)integrability and even for the peculiar shape of the spectral curves. This adds to the older questions of ambiguity of $W$-representations and of possibility of selecting operators $\hat{\cal W}$ belonging to the $w_\infty$ algebra.  We hope that many of these questions will attract interest and will be addressed and answered in the near future.

\section*{Acknowledgements}

We are grateful to V. Mishnyakov and A. Popolitov for useful discussions. We are also indebted to the referee of this paper for stimulating questions. This work was supported by the Russian Science Foundation (Grant No.21-12-00400).

\end{document}